# Search for new pulsating O VI central stars of planetary nebulae *

Roberto Silvotti[1], Corrado Bartolini[1], and Letizia Stanghellini[2]

(1) Dipartimento di Astronomia, Università di Bologna, via Zamboni 33, 40126 Bologna, Italy
(2) Osservatorio Astronomico di Bologna, via Zamboni 33, 40126 Bologna, Italy

**Abstract.** A photoelectric monitoring program has been applied, during the last four years, to five central stars of planetary nebulae (PNNs) with strong O VI $\lambda$3811–34 Å emission. NGC 6905 and, marginally, NGC 2452, show intrinsic luminosity variations, while NGC 7026, IC 2003 and NGC 1501 have constant luminosity within a few mmag. Photometric data have been analyzed with the best available packages for power–spectra reductions. Both pulsators have periods and physical characteristics well encompassed by the theoretical pulsational models relative to these stars.

## 1. Introduction

In the evolutionary transition zone between the end of the constant luminosity phase – at the tip of the asymptotic giant branch (AGB) – and the beginning of the white dwarf (WD) cooling phase, a group of very hot stars shows luminosity fluctuations due to the presence of nonradial g-mode pulsations. Some of them are variable PN nuclei (PNNV), the remaining are PG 1159-type objects (called also DOV).

The hypothesis that the PNNV are the progenitors of the PG 1159 stars is strongly suggested by the common spectroscopic properties, the increasing surface gravities and the decreasing pulsation periods from PNNV to DOV. The typical values of the pulsation periods are 10–35 min for the PNNV and 5–15 minutes for the DOV stars. The common spectroscopic properties reflect a similar photospheric composition, high C/O abundances, relatively low He abundances ($\sim$30% by mass) and extreme H deficiency

Send offprint requests to: silvotti@astbo3.bo.astro.it

* Based on observations obtained at the Loiano Observatory, Italy, and at the European Southern Observatory, La Silla, Chile.

(Werner 1992). The recent discovery that the hot pulsating WD RXJ 2117.1+3412, with the highest pulsation periods among DOV stars (Vauclair et al. 1993), has a large low-surface-brightness PN (Appleton et al. 1993), confirms the evolutionary link between PNNV and DOV stars.

The photospheric abundances of the PNNV and the PG 1159 stars agree with the C/O $\kappa - \gamma$ mechanism proposed by Starrfield et al. (1984) for keeping the pulsational equilibrium in time. Moreover, high oxygen abundance is required to sustain pulsations in the hottest stars, with effective temperature beyond 100 000 K. Such high oxygen abundance is present in all the known PNNV and is attested by the strong $\lambda$ 3811–3834 Å O VI emission doublet. The $\kappa - \gamma$ mechanism is active in layers close to the stellar surface, at temperatures of the order of $10^6$ K, and is highly sensitive to the chemical composition. Small amounts of He or H in the C/O partial ionization zone are sufficient to inhibit the delicate pulsational mechanism (Stanghellini et al. 1991). Some helium and hydrogen can exist only in a thin surface layer. The high sensitivity to the chemical composition of the pulsation mechanism may explain why not all the O VI PNNs do pulsate.

The presence of trapped modes, which are preferred with respect to the other pulsation modes because of their lower kinetic energy, gives an explanation to the negative secular change of the pulsation periods, found in the DOV stars PG 1159-035 (Winget et al. 1991) and PG 1707+427 (Grauer et al. 1992). In fact, trapped modes are more sensitive to the envelope contraction than to the cooling of the star.

An updated picture of the pulsating/nonpulsating O VI PNNs is shown in a recent review by Bond et al. (1993), while the properties of the PG 1159 stars are reviewed by Werner (1992), and, more recently, an overall study on the physical properties of O VI PNNs and their nebulae has been completed by Stanghellini (1995).

Currently there are 18 known O VI PNNs: 7 of them are pulsating, 4 are nonpulsating, for the remaining the variability is not clear or totally unknown (Bond et al.

Table 1. Log of observations and results

| PNN | V | JD | $d^{hours}$ | filter | $L_{cs}^{mmag}$ | $L_{cc}^{mmag}$ | $n^{mmag}$ | $\Pi^{min}$ |
|---|---|---|---|---|---|---|---|---|
| NGC 1501 | 14.39 | 48510 | 2.0 | B | 5 | 3 | 2.3 | |
| NGC 2452 | >17 | 48918 | 1.5 | U | 11 | 9 | 3.9 | 3.6, (~ 40) |
| NGC 6905 | 15.7 | 48506 | 3.9 | B | 5 | 3 | 1.2 | (15.0, 23.5) |
| | | 48509 | 2.9 | B | 7 | 4 | 1.5 | 14.5, (23.5) |
| | | 48518 | 3.2 | U | 10 | 7 | 2.1 | 14.7 |
| NGC 7026 | 14.20 | 48507 | 1.1 | B | 6 | 6 | 2.5 | |
| | | 48508 | 3.9 | B | 5 | 5 | 0.7 | |
| | | 48510 | 3.0 | B | 6 | 4 | 0.8 | |
| | | 48516 | 0.5 | U | 8 | 6 | 3.7 | |
| | | 49611 | 3.0 | U | 7 | 3 | 1.8 | |
| IC 2003 | 15.0 | 48506 | 2.3 | B | 8 | 7 | 1.2 | |
| | | 48508 | 2.2 | B | 3 | 2 | 0.8 | |
| | | 48509 | 2.9 | B | 3 | 2 | 0.9 | |
| | | 48516 | 2.7 | U | 6 | 5 | 2.5 | |

1993, Bond 1994). The aim of this work is to investigate the variability of these latter O VI PNNs.

## 2. Observations and data analysis

The targets have been chosen from the set of the O VI PNNs, favoring those whose variability has not yet been investigated. Four northern PNNs, NGC 1501, NGC 6905, NGC 7026 and IC 2003, have been observed by means of the two head photometer mounted on the 1.5 m reflector at the Loiano Observatory, Italy. The only southern PNN of our sample, NGC 2452, has been observed from La Silla, Chile, with the monochannel photometer of the 1.0 m ESO telescope. The instrumentation and the observational techniques used here are described in some detail by Silvotti et al. (1995, and references therein). The observing strategy in this paper differs from that of Silvotti et al. (1995) only by the filters used (Johnson U and B, see Table 1) and in the sampling times, equal to 10 s for NGC 2452, and between 2 and 5 s for the northern PNNs. After a preliminary analysis with the original time resolution, the data of the northern PNNs have been analyzed after having been summed to an effective integration time of 20 s.

The target list is given in Table 1: the apparent Johnson V magnitudes of the central stars (CS), taken from Acker et al. (1992), are in column (2); the julian day of each observation is in column (3); the duration of each observation is in column (4); the upper limits to the pulsational amplitude and the noise of the Fourier transforms, all in millimagnitudes, are in column (6), (7) and (8). The pulsational periods found from our data analysis are in column (9), in brackets when uncertain.

The limiting amplitudes in column (6) and (7) have been obtained with the CLEAN algorithm (Roberts et al. 1987). The number in column (6) is the amplitude of the highest peak in the "clean spectrum", whereas the number in column (7) is the amplitude of the highest "clean component". The noise in column (8) is the average of all the peaks present in each "clean spectrum". This number gives an estimate of the quality of the results, depending on both the quality and the length of the measurements. Both the limiting amplitudes and the noises are valid in a period range between 40 s (20 s for NGC 2452) and the half of the duration of each run. For observations longer than 2.2 hours, the period range is 40–4000 s. The limit of 4000 s has been posed because, in the extreme low frequency region of the power spectra, we can find some peaks introduced probably by spurious effects, like sky variations and/or inaccuracies in the extinction corrections applied to the target and the comparison star. To reveal pulsations with periods longer than about 4000 s, we would need observations on a quite longer time-scale.

The data have been reduced by means of the program BIC, written by R. Silvotti, which makes dead-time correction, sky background subtraction, extinction correction and reduction to baricentric julian days, by means of the Stumpff's (1980) algorithms.

The analysis of the data have been performed using different codes, based on the method of Deeming (1975) modified by Kurtz (1985), the method of Lomb (1976) and Scargle (1982) for unevenly sampled data (see also Press et al. 1992, par.13.8), and the CLEAN method (Roberts et al. 1987).

## 3. Results and discussion

**NGC 1501**: the power spectra do not show prominent peaks. The amplitudes of the highest clean components, at about 7.0, 9.6 and 12.3 mHz, do not exceed 3.3 mmag (Fig. 1). Bond & Ciardullo (1991) have detected pulsations in the nucleus of NGC 1501, with typical periods around 24 min. We found that, in that region of the spectrum, the amplitudes of the clean components do not exceed 1.8 mmag.

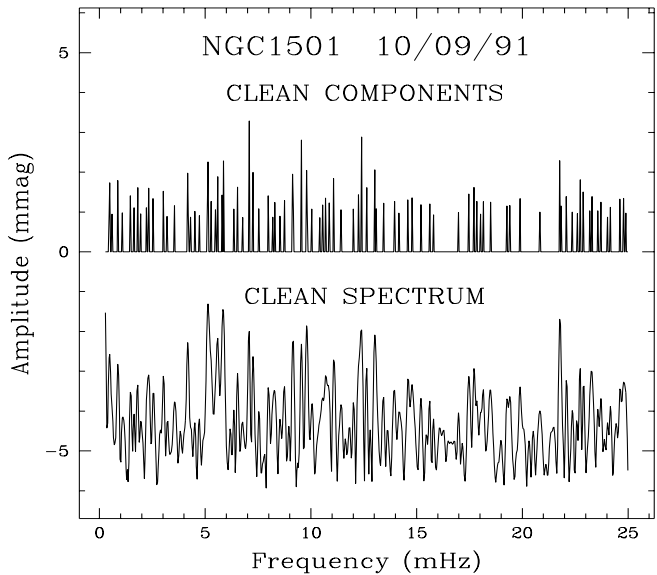

**Fig. 1.** Power spectra of the nucleus of NGC 1501 in the period range 40–3600 s

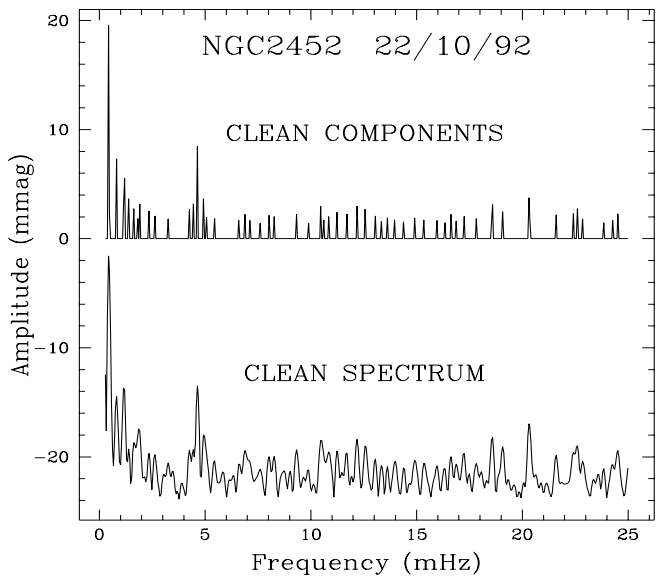

**Fig. 2.** Power spectra of the nucleus of NGC 2452 in the period range 40–3500 s. The highest peak at about 40 min (0.42 mHz) needs longer observations to be confirmed. More convincing is the peak at about 215 s (4.6 mHz)

**NGC 2452**: the peak at 4.6 mHz has a clean component amplitude (clean amplitude) of 8.5 mmag (Fig. 2). The length of the observation, 1.5 hours, is quite large respect to the corresponding period of $(217 \pm 9)$ s, and therefore the periodicity that we found should be real. A second peak, with a period of about 40 min and a clean amplitude of 19.5 mmag, is visible in the light curve, but it is more uncertain. The sky conditions at the time of the observation of the nucleus of NGC V 2452 were good (seeing of ~1 arcsec), but the faintness of the star, which contributes substantially to the noise of the Fourier transform, requires new observations to confirm its variability. A new observing-time request will be submitted to ESO for this purpose.

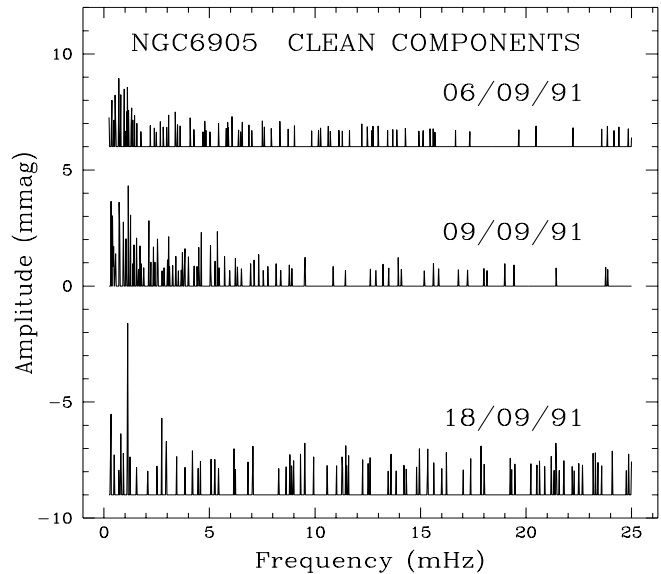

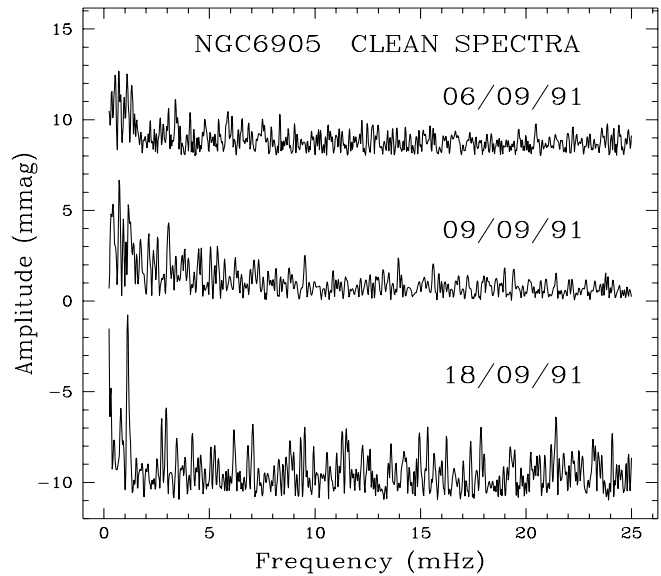

**Fig. 3.** Power spectra of the nucleus of NGC 6905 in the period range 40–4000 s. The peak at about 885 s (1.1 mHz) is more prominent in the U observation of the 18/09/91, but is also present in the B observation of the 09/09/91

**NGC 6905**: the peak at about 1.1 mHz is clearly present in both power spectra relative to the observations of September 9 and 18, 1991 (Fig. 3). In the first observation, with the B filter, the clean amplitude is equal to 4.3 mmag. In the second one, with the U filter, the amplitude is higher and reaches 7.4 mmag. There are two advantages and one disadvantage using the U filter instead of the B filter. First, the nonradial pulsations show their maximum amplitude in the UV band, because the luminosity fluctuations are almost completely due to the temperature variations, whereas the surface variations are not important. Second, the U light is less affected by the contribution of the nebula. The disadvantage is obvious: worse S/N ratio, and therefore more noise in the Fourier transform, as we can see from column (8) of Table 1. In the power spectrum relative to the observation of September 6, 1991, it is more difficult to distinguish the peak at 1.1 mHz. Several peaks are distributed between 0.2 and 1.6 mHz; the highest, with a period of about 24 min (0.7 mHz), has a clean amplitude of 3 mmag and is also present the 09/09/91. The variability of the CS of NGC 6905 was claimed by Bond & Ciardullo (1991). They found typical periods of 16 min. From our data, the mean pulsation periods is at $(14.7 \pm 1.1)$ min.

**NGC 7026**: the power spectra do not show any common peak in different nights (Fig. 4). The peaks at the extreme left side of the frequency range, between 0.2 and 0.25 mHz, are probably due to spurious effects, like sky variations and/or wrong extinction corrections for the target and the comparison star. If we consider the best observations, with the lowest noise, made in 1991 Sept 8 and 10, and 1994 Sept 15, the clean amplitudes are less than 2.2 mmag (B band) or 2.8 mmag (U band) in the frequency range 0.3–25 mHz. Such limits are reduced to 1.7 mmag in the range 1–25 mHz with both the B and U filters.

**IC 2003**: the amplitudes of the highest clean components are less than 1.5 mmag (Sept 8, 1991) or 2.1 mmag (Sept 9, 1991) in the frequency range 0.25–25 mHz, using the B filter (Fig. 5). The limiting amplitude is much higher, 7.3 mmag in B (Sept 6, 1991) for the same frequency range. In fact, the data of Sept 6 are of lower quality with respect to the other nights. Therefore we think that the two peaks at about 0.4 mHz are not real. Finally, in the U observation of Sept 16, 1991, the limiting amplitude is equal to 4.7 mmag in the range 0.25–25 mHz. There are little coincidences between the peaks in the power spectra of different nights: the peak at about 10.1 mHz is present in both spectra of Sept 9 and 16, 1991, with a clean amplitude of 1.2 mmag (B band) and 3.7 mmag (U band). Some coincidences are also present in the low frequency region, between 0.7 and 1.7 mHz.

For the stars with more than one observation, NGC 6905, NGC 7026 and IC 2003, further analyses have been done, after having added the data of close different nights in a single file. The results confirm those obtained from single nights.

In Fig. 6 we show how the two pulsating stars found from our monitoring compare with the nonradial pulsational models from Stanghellini et al. (1991). In the Figure, we plot the period–temperature loci of the two stars superimposed to the period–temperature ranges allowed for PNNs pulsations. The dots represent the first periods detected in this paper, and the arrows joint these points to the secondary (less probable) periods. Temperatures of the stars are from Stanghellini (1995). We can see from the figure that the observed periods are within the domine of the theoretical ones, giving us additional confidence in our measurements. A wider sample of O VI central stars in this diagram has been shown in Stanghellini et al. (1995).

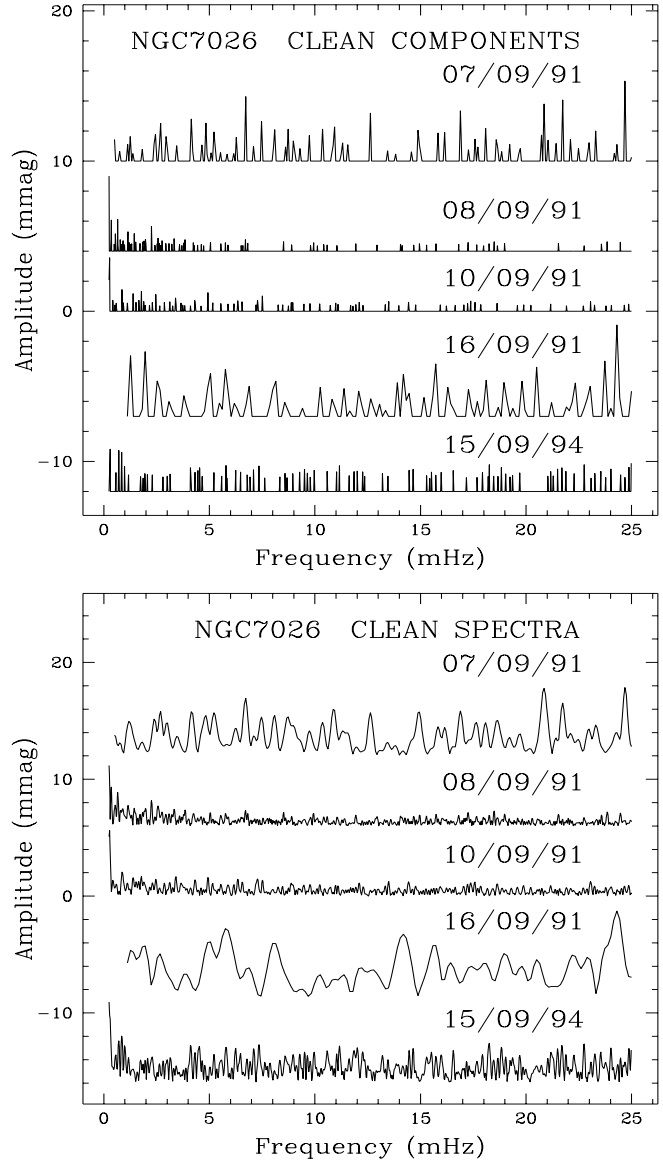

**Fig. 4.** Power spectra of the nucleus of NGC 7026 in the period range 40–4000 s. There are no common peaks in the different nights

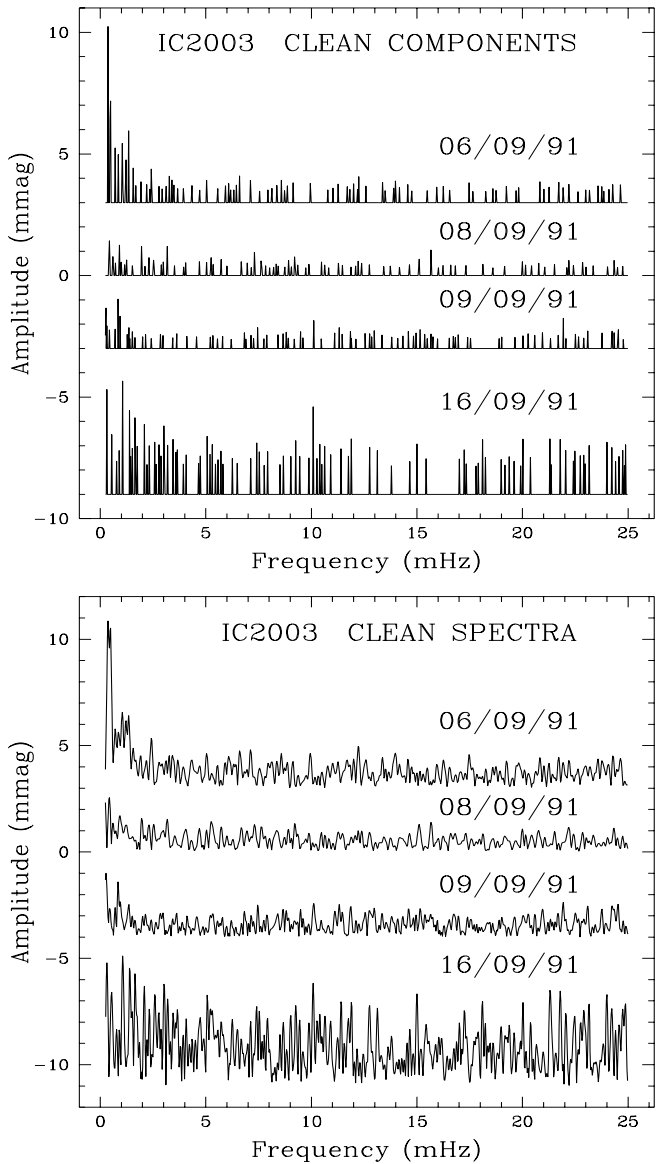

Fig. 5. Power spectra of the nucleus of IC 2003 in the period range 40–4000 s

## 4. Summary

Following the results of our observations, the nuclei of NGC 7026 and IC 2003 do not pulsate with amplitudes higher respectively than 3 mmag (NGC 7026) and 3 or 5 mmag (IC 2003, B or U filter), in a period range between 40 s and more than 3300 s. New observations of the nucleus of IC 2003, with a longer time-scale, could confirm its nonvariability. For the CS of NGC 1501, observed just once, the limiting amplitude is equal to 4 mmag. The nucleus of NGC 6905 does pulsate, according with Bond & Ciardullo (1991). The main pulsational period registered is at $(14.7 \pm 1.1)$ min. Another periodicity of about 24 min is not as sure. From the single observation of NGC 2452, its CS seems to be pulsating. A periodicity of $(217 \pm 9)$ s

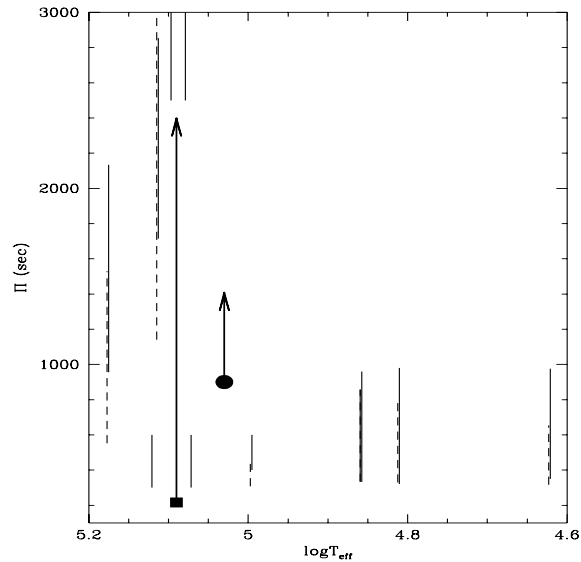

Fig. 6. Period–effective temperature diagram for theoretical modes l=1 (solid lines) and l=2 (broken lines), from Stanghellini et al. (1991). The locations of newly detected periods are shown, the square being NGC 2452, the circle being NGC 6905

has been registered, another one of about 40 min is less certain. New observations of the nucleus of NGC 2452 are needed to confirm its variability.

*Acknowledgements.* This research was partially supported by the Italian "Ministero per l'Università e la Ricerca Scientifica e Tecnologica" (MURST). We gladly thank Howard Bond for giving us updates of pulsational periods of evolved stars, and Joseph Lehár for making the CLEAN algorithm codes available to us.

## References

Acker A., Ochsenbein F., Stenholm B., et al., 1992, Strasbourg-ESO Catalogue of Galactic Planetary Nebulae, (ESO: Garching)
Appleton P.N., Kawaler S.D., Eitter J.J., 1993, AJ 106, 1973
Bond H.E., 1994, private communication
Bond H.E., Ciardullo R., 1990, in Confrontation between Stellar Pulsation and Evolution, eds. C. Cacciari and G. Clementini, PASP Conf. Series, vol.11, p.529
Bond H.E., Ciardullo R., Kawaler D.E., 1993, Acta Astron. 43, 425
Deeming T.J., 1975, Ap&SS 36, 137
Grauer A.D., Green R.F., Liebert J., 1992, ApJ 399, 686
Kurtz D.W., 1985, MNRAS 213, 773
Lomb N.R., 1976, Ap&SS 39, 447
Press W.H., Teukolsky S.A., Vetterling W.T., Flannery B.P., 1992, Numerical Recipes, (Cambridge Univ. Press: Cambridge)
Roberts D.H., Lehár J., Dreher J.W., 1987, AJ 93, 968
Scargle J.D., 1982, ApJ 263, 835
Silvotti R., Bartolini C., Cosentino G., Guarnieri A., Piccioni A., 1995, A&A, submitted

Stanghellini L., Cox A.N., Starrfield S., 1991, ApJ 383, 766

Stanghellini L., 1995, Ph. D. Thesis, University of Illinois

Stanghellini L., Kaler J. B., Shaw R. A., & di Serego Alighieri S., 1995, A&A, in press

Starrfield S., Cox A.N., Kidman R.B., Pesnell W.D., 1984, ApJ 281, 800

Stumpff P., 1980, A&AS 41, 1

Vauclair G., Belmonte J.A., Pfeiffer B., et al., 1993, A&A 267, L35

Werner K., 1992, in Atmospheres of Early-type Stars, eds. U. Heber and C.S. Jeffery, Lecture Notes in Physics 401, (Springer-Verlag: Heidelberg), p.273

Winget D.E., Nather R.E., Clemens J.C. , et al., 1991, ApJ 378, 326